\documentclass[%
 sor,
 jor,
 amsmath,amssymb,
 reprint,%
]{revtex4-1}

\usepackage{graphicx}
\usepackage{dcolumn}
\usepackage{bm}
\usepackage{color}
\usepackage{amstext}
\usepackage{amssymb}
\usepackage{graphicx}
\usepackage{amsmath}
\usepackage{bm}
\usepackage[mathscr]{euscript}
\usepackage{amsfonts}

\usepackage{tcolorbox}
\newtcolorbox{mybox}[1]{colback=red!5!white,colframe=red!75!black,fonttitle=\bfseries,title=#1}

   \newcommand{\bom}{\mbox{\boldmath $\omega$}}

\newcommand{\tn}{\textnormal}

   \newcommand{\twothirds}{\textstyle{\frac{2}{3}} \displaystyle }

\newcommand\be{\begin{equation}\,}
\newcommand\ee{\end{equation}\,}

\makeatletter
\def\@makefnmark{\hbox{\@textsuperscript{\normalfont\@thefnmark}}}
\makeatother

\begin{document}

\preprint{AIP/123-QED}

\title[Singularities in Fluid Mechanics]
{Singularities in Fluid Mechanics}

\author{H. K. Moffatt}

\affiliation{Department of Applied Mathematics and Theoretical Physics, \\
University of Cambridge, Wilberforce Road, Cambridge CB3 0WA, UK}

%

\date{\today}

\begin{abstract}

\noindent Singularities of the Navier-Stokes equations occur when some derivative of the velocity field is infinite at any point of a field of flow (or, in an evolving flow, becomes infinite at any point within a finite time).  Such singularities can be mathematical (as e.g.~in two-dimensional flow near a sharp corner, or the collapse of a M\"{o}bius-strip soap film onto a wire boundary) in which case they can be ‘resolved’ by refining the geometrical description; or they can be physical (as e.g.~in the case of cusp singularities at a fluid/fluid interface) in which case resolution of the singularity involves incorporation of additional physical effects; these examples will be briefly reviewed. The `finite-time singularity problem' for the Navier-Stokes equations will then be discussed and a recently developed analytical approach will be presented; here it will be shown that, even when viscous vortex reconnection is taken into account, there is indeed a physical singularity, in that, at sufficiently high Reynolds number, vorticity can be amplified by an arbitrarily large factor in an extremely small point-neighbourhood within a finite time, and this behaviour is not resolved by viscosity.   Similarities with the soap-film-collapse and free-surface-cusping problems are noted in the concluding section, and the implications for turbulence are considered.\\
%

\end{abstract}

\pacs{Valid PACS appear here}
\keywords{Suggested keywords}
\maketitle

\section{Introduction}
This paper is an update of the Otto Laporte lecture, delivered at the APS/DFD meeting in Atlanta, GA.,\,18 November 2018. Its main purpose was to describe an approach to the finite-time singularity problem for the Navier-Stokes equations, as recently developed in collaboration with Yoshifumi Kimura, Nagoya University (Moffatt \& Kimura 2019a,b).   However, by way of warm-up to this famous unsolved problem, I first described three situations where singularities of various kinds appear, but always at the boundary of the fluid domain.  These problems are briefly reviewed in \S\S II-IV below; and our approach to the central Navier-Stokes problem is summarised in \S\S V and VI.

\section {A corner singularity}\label{sec_corner_singularities}
\begin{figure}
(a)\includegraphics[width=.7\linewidth]{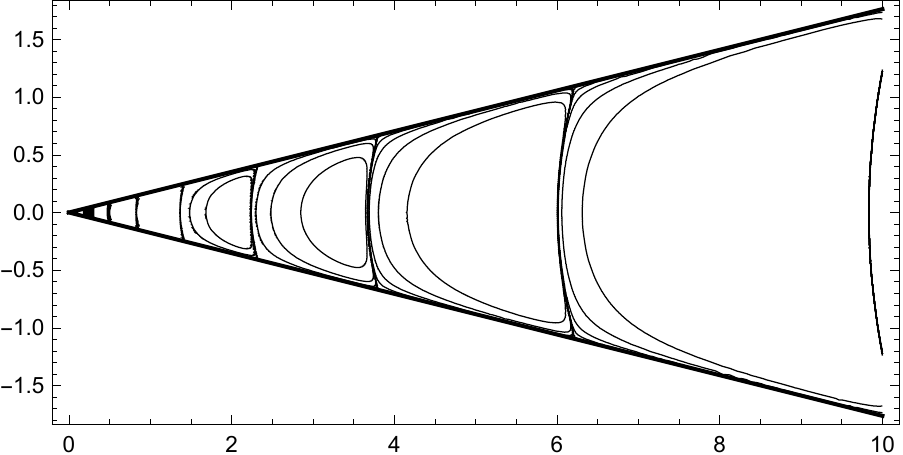}\\
\vskip 2mm
(b)\includegraphics[width=0.45\textwidth, trim=30mm 30mm 30mm 0mm]{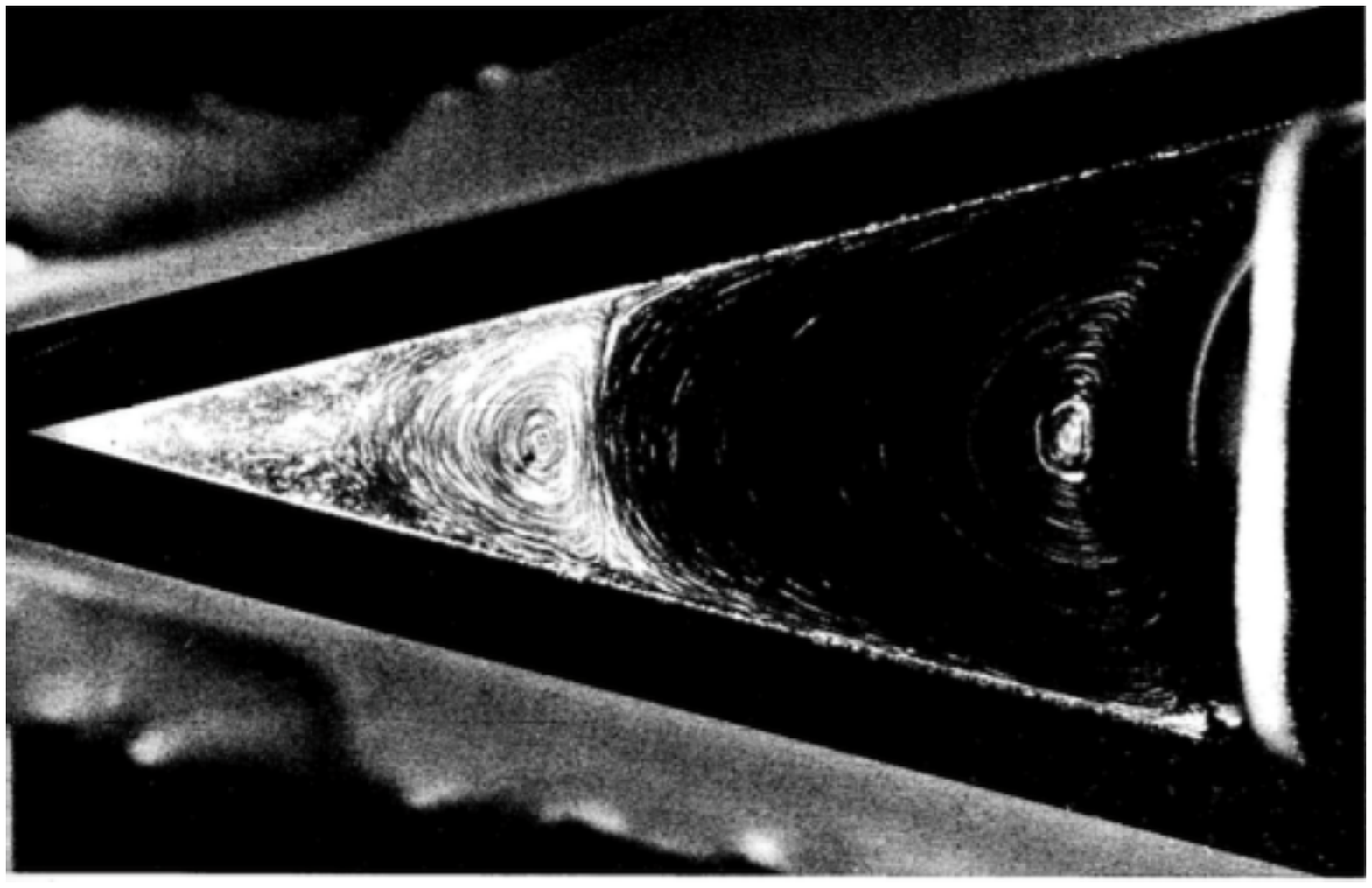}\\  
(c)\includegraphics[width=0.40\textwidth, trim=30mm 80mm 0mm 80mm]{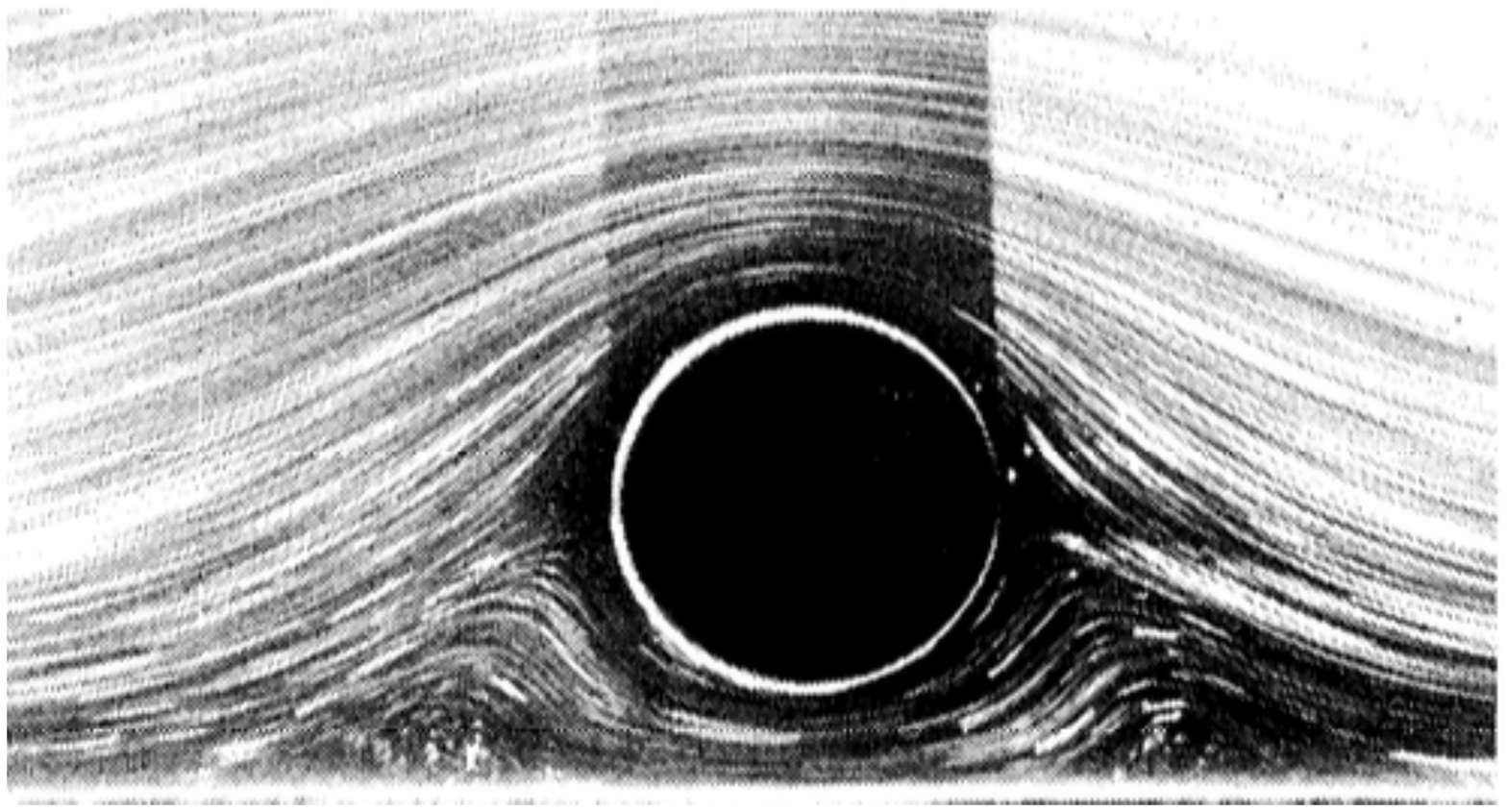}\\  (d)\includegraphics[width=0.40\textwidth, trim=30mm 80mm 0mm 80mm]{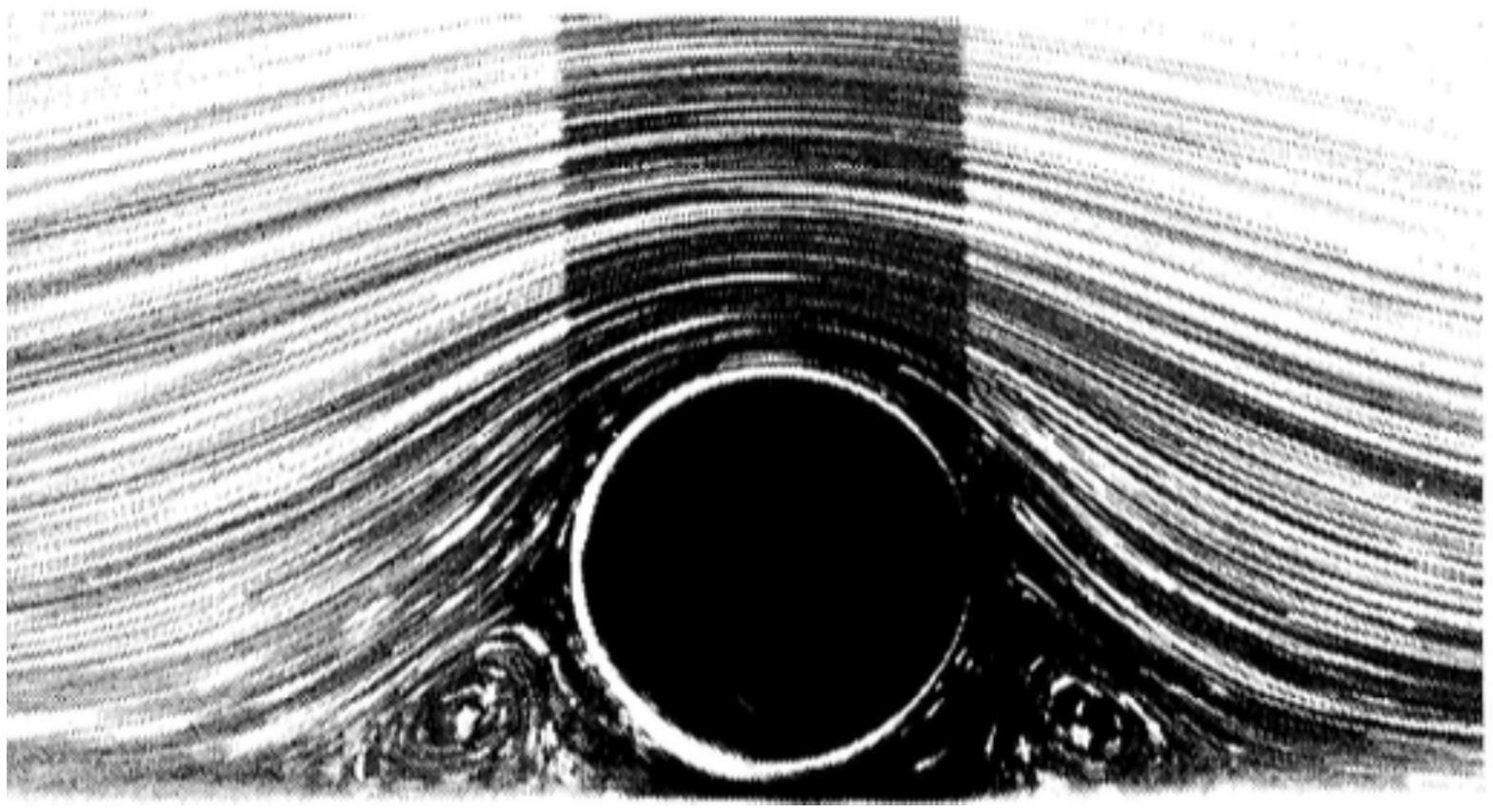}
        \caption{(a) Corner eddies driven by antisymmetric excitation far from the corner.  (b) This flow is driven by a rotating cylinder; a primary eddy and a secondary eddy are clearly visible; this is a Stokes flow, in which flow separation and reattachment are evident. (c) Similar Stokes separation in shear flow over a cylinder near a plane boundary; (d) the same with the cylinder very near the plane. [Photos from  Taneda 1979, \copyright 1979 The Physical Society of Japan.]}
        \label{Fig_corner_eddies}
 \end{figure}

 Two-dimensional flow near a sharp corner exhibits a curious singularity that has been the subject of many investigations dating back to Lord Rayleigh (1920), Dean \& Montagnon (1949), and Moffatt (1964). Assuming no-slip on both walls $\theta=\pm\alpha$, the local Reynolds number is very small near the corner, and the stream-function $\psi(r,\theta)$ satisfies the biharmonic equation $\nabla ^{4} \psi =0.$  The relevant solution has the asymptotic form $\psi \sim r^{\lambda} f(\theta)$, and it turns out that, for $2\alpha \lesssim 147^{\tn{o}}$, the exponent $\lambda$ is complex: $\lambda=p+\textnormal{i}q$, say, with $p>1$ and $q\ne 0$, both functions of the angle $\alpha$.  This much was recognised by Dean \& Montagnon; what is interesting however is the resulting structure of the solution:  there is an infinite geometric sequence of eddies, decreasing rapidly in intensity, as the corner is approached, as shown in 
 Fig.~\ref{Fig_corner_eddies}(a).
 This singular behaviour is associated with the assumed singularity of the boundary, the curvature being infinite at the corner. No corner is perfectly sharp in reality, and if  it is rounded off, the singularity is removed and the number of eddies is finite; however the phenomenon of flow separation is still apparent, even although this is a Stokes flow.
 
 This phenomenon was realised experimentally by Taneda (1979)  (see the striking photograph in Fig.~\ref{Fig_corner_eddies}(b) reproduced in Van Dyke's 1982 famous \emph{Album of Fluid Motion}). The phenomenon is quite universal, in that this Stokes separation occurs whenever fluid is forced into a corner or rapidly converging region. For example, a similar sequence of eddies exists in shear flow over a cylinder near a plane boundary 
 (Figure \ref{Fig_corner_eddies}(c,d), also from Taneda 1979).  If there is a small gap between the cylinder and the plane, then there is a small `leakage' through the gap, and the eddies (again finite in number) are attached alternately to the cylinder and the plane, the number of these eddies  increasing without limit (in principle) as the gap decreases. 
 Many variants may be found in the book of Shankar (2007), which is almost wholly devoted to this type of phenomenon. 
 
 \section{A twist singularity}\label{sec_twist_singularity}
   \begin{figure}
\vskip 2 mm
(a)\includegraphics[width=0.5\textwidth, trim=20mm 100mm 0mm 110mm]{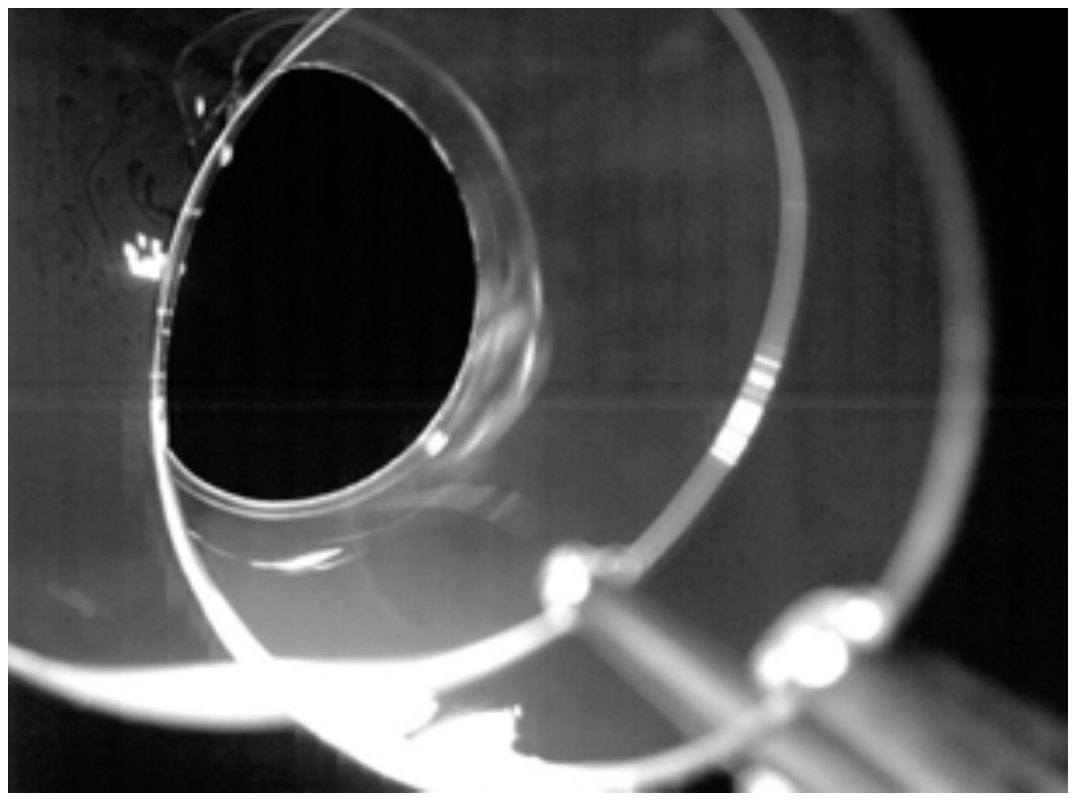}\\
\vskip 2mm
(b)\includegraphics[width=0.3\textwidth, trim=0mm 0mm 0mm 0mm]{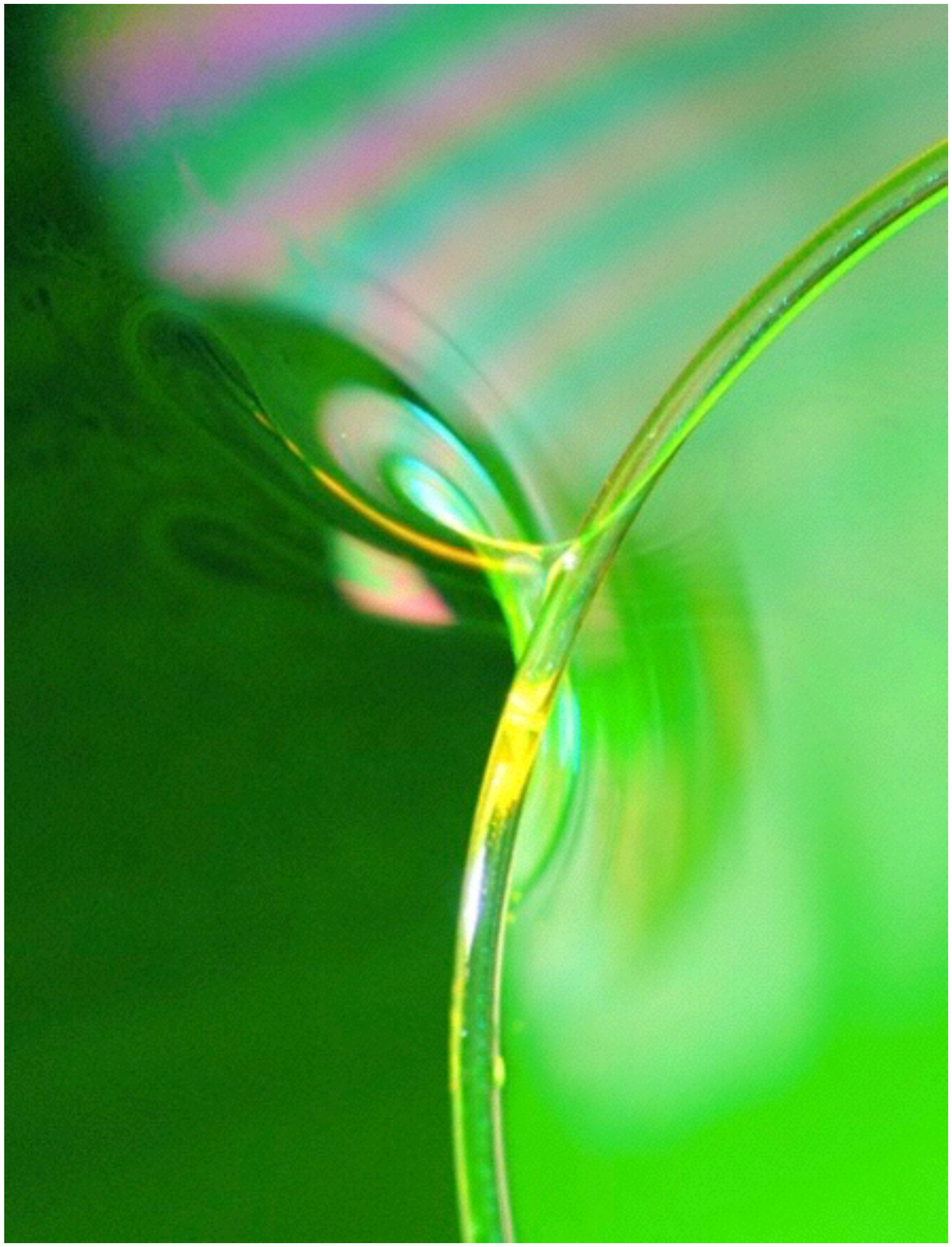}
        \caption{(a) A one-sided soap-film in the form of a M\"{o}bius strip of minimal area is supported by a wire boundary, which may be slowly untwisted  until instability causes the hole to collapse suddenly to the boundary; (b) the twist singularity, observed by reflection from the twisted surface immediately after the collapse; at this stage the surface is two-sided. [From Goldstein et al. 2010]}
        \label{Fig_twist_singularity}
 \end{figure}
Fig.~\ref{Fig_twist_singularity}(a) shows a soap film suspended on an initially circular wire that is twisted and folded back on itself in such a way as to be the boundary of a M\"{o}bius strip; this is a one-sided soap-film of minimal area.  When the wire is very slowly untwisted, at a critical moment the hole in the 
M\"{o}bius strip collapses in a matter of milliseconds to a singularity on the wire boundary, which shows up optically as a very localised twist in the soap film itself (Fig.~\ref{Fig_twist_singularity}(b)).  After this collapse, the film is two-sided, topologically a disc!  Thus a topological transition is induced at the moment of collapse.  

This phenomenon has been investigated by Goldstein et al.~(2010), and in several subsequent papers (Goldstein, Moffatt \& Pesci 2012, Goldstein et al. 2014, Moffatt, Goldstein \& Pesci 2016).  In order to resolve the singularity at the instant when the hole disappears, it is necessary to take account of the finite cross-sectional radius $\delta$ of the wire (just as it is necessary to take account of the finite radius of curvature  at the `corner' in the above corner-flow problem). The Plateau boundary where the projection of the soap film meets the wire is a closed curve on the surface of the wire, which, just before the  jump, is  doubly linked with the axis of the wire, with (here) a left-handed localised $2\pi$-twist round the wire in the immediate neighbourhood of the incipient singularity.  Just after the jump, there is a right-handed $2\pi$-twist round the wire at the same location (this can just be discerned in 
Fig.~\ref{Fig_twist_singularity}(b)). It is the viscous stress on the wire and in its immediate neighbourhood that is responsible for this change from left-handed to right-handed twist.  The net effect is that the double linkage characteristic of the
M\"{o}bius strip is replaced by the zero-linkage of Plateau boundary and wire characteristic of a two-sided soap film with disc topology.

 The kinetic energy of the collapsing film is increasingly concentrated at  the singularity where it is very rapidly dissipated by viscosity. The  energy dissipated in this way must obviously be equal to the net loss of surface energy of the soap film (proportional to its area) during the collapse process.

 \section{A free-surface cusp singularity}\label{free-surface_cusp}
\begin{figure}
\vskip 2 mm
(a)\includegraphics[width=0.38\textwidth, trim=0mm 180mm 0mm 0mm]{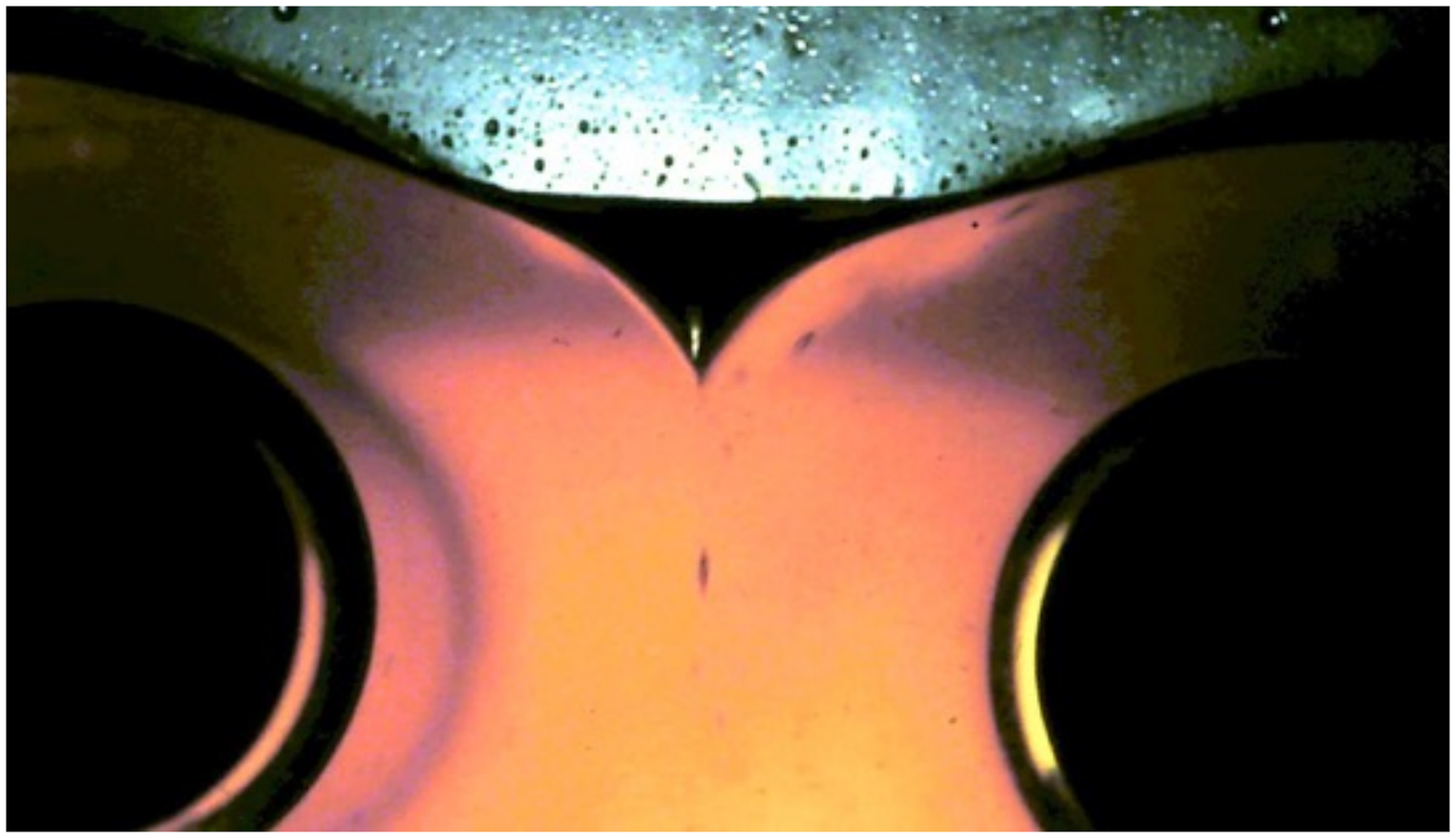}\\
\vskip 2mm
(b)\includegraphics[width=0.38\textwidth, trim=0mm 180mm 0mm 0mm]{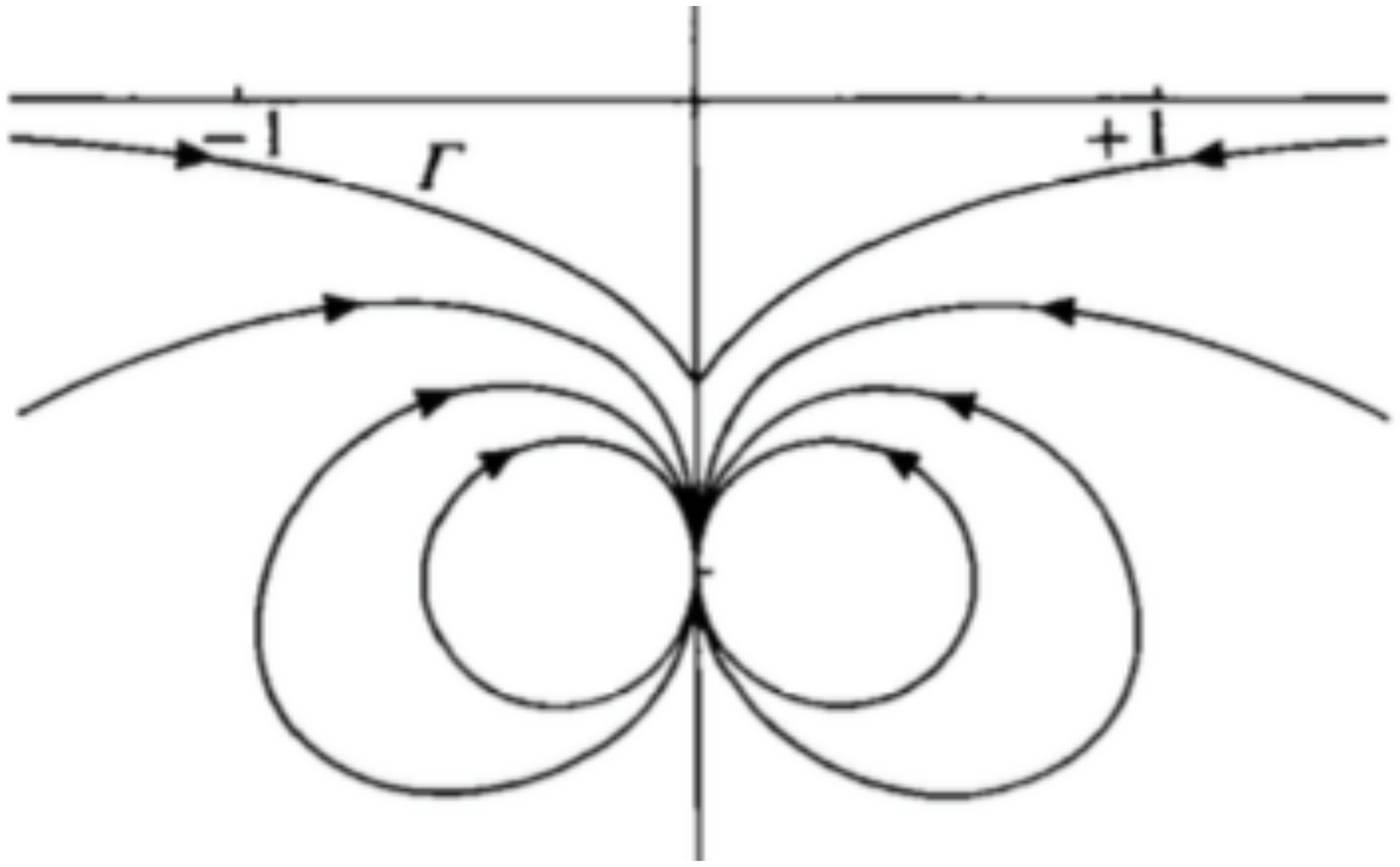}
        \caption{(a) A cusp at the free surface  of a viscous liquid induced by sub-surface counter-rotating cylinders;  (b) flow modelled by a vortex dipole at depth $d$ below the undisturbed free surface; the cusp appears at depth $\twothirds d$. [From Jeong \& Moffatt 1992.]}
        \label{Fig_cusp}
 \end{figure}

Fig.~\ref{Fig_cusp}(a) shows a cusp singularity on the free surface of a viscous fluid (here golden syrup), the flow being driven by two sub-surface cylinders counter-rotating in the sense that makes the flow on the free surface converge towards the plane of symmetry. The streak on the centre-line is the remnant of a small  air bubble that has been drawn in through the cusp, an important consequence of the cusping phenomenon. Fig.~\ref{Fig_cusp}(b) shows an idealised configuration for which mathematical analysis is possible:  a two-dimensional vortex dipole of strength $\alpha$ (representing the effect of the rotating cylinders) is placed at depth $d$ below the undisturbed surface; the fluid has  surface tension $\gamma$, and viscosity $\mu$ sufficiently large that Stokes flow may be assumed.  The only dimensionless parameter governing the flow is then the capillary number $C=\mu\alpha/d^{2}\gamma$.

This two-dimensional problem admits solution via complex variable techniques involving conformal transformation of the fluid domain (Jeong \& Moffatt 1992). The surface is indeed drawn down, the minimum being at depth $\sim 2d/3$, i.e. well above the imposed vortex-dipole singularity.  The radius of curvature $R$ at this minimum point is given by the exact formula
\be
R/d=(256/3)\exp [-32\pi C]\,,
\ee
and is  extremely small when $C=\tn{O}(1)$.  Indeed, if we adopt a `level playing field', i.e.~$C=1$ as between the competing effects of viscosity and surface tension, then $R/d \approx 1.9 \times 10^{-42}$, surely the smallest $\tn{O}(1)$ number to emerge from any fluid dynamical problem!  This is  perfectly regular from a purely mathematical point of view, but may be described as a `physical singularity' which exists despite the smoothing effect of surface tension.

Resolution of this singularity was proposed by Eggers (2001), who recognised that the build-up of air pressure in the cusp region will lead to entrainment of bubbles through the cusp, as observed in practice.  The effect is familiar even in high Reynolds number situations:  when a weak stream of water flows from a tap into a deep bath of water,  a circular cusp tends to form where the stream makes impact with the free surface.  As the flow rate is increased, a critical stage is reached at which an instability in the cusp region leads to entrainment of bubbles through this cusp, with audible effect.  This type of cusp-entrainment of one fluid into another is presumably a key process at fundamental level in the mixing of two otherwise immiscible fluids. 

\section{Finite-time singularity problem}
The well-known Clay millennium prize question is essentially this:  can any initially smooth velocity field of finite energy in an incompressible fluid become singular at finite time under Navier-Stokes evolution?  The question dates back to Leray's (1934) seminal paper, and has attracted intense investigation over recent decades. The problem is important not only in its own right in relation to the regularity of the Navier-Stokes equations. It also has important application in relation to the  smallest scales of turbulence at which energy is dissipated;  here, the question is: how exactly is it that, with a turbulent vorticity field $\bom({\bf x}, t)$, the rate of dissipation of energy $\Phi=\nu \langle {\bom}^2 \rangle$  can remain finite in the limit of vanishing kinematic viscosity $\nu$?  This requires that the enstrophy $\langle {\bom}^2 \rangle$ must become infinite, so that singularities in $\bom({\bf x}, t)$ must be present in the limit at each time instant $t$.  What then is the nature of these singularities?

There is a huge background literature to the finite-time singularity problem, as it has come to be called, and we can mention only some highlights here.  It is widely believed that the configuration most likely to lead to a sing\-ularity consists of two interacting non-parallel vortex tubes: this belief is based on the work of Beale, Kato \& Majda (1984) who showed that a necessary condition for a  singularity at the finite time $t^*$ is that $\int^{t} \tn {max}|\bom|\tn{d}t$ must diverge as $t\rightarrow t^*$. Moreover, Constantin, Fefferman \& Majda (1996)  showed that, at least for the Euler equations,  the direction  of the vorticity must be indeterminate in the limit as the singularity is approached.  A further paper of great significance is that of de Waele \& Aarts (1994), who considered many different initial conditions for interacting vortices, and found that ``on their way to the collision, the two vortices form a pyramid structure which is independent of the initial conditions \dots ; [\emph{the vortices}] follow a universal route for all kinds of initial arrangements''.  These results place the search for a singularity firmly within the scenario of interacting vortices, as originally proposed  by Siggia (1985) and Siggia \& Pumir (1985) , who  had earlier considered such interactions and who 
asserted that ``A local model should exist which would allow one to understand the stage of rapid stretching in terms of simpler processes".  It is just such a local model that we have sought to develop, as described below.
 \vskip 2mm
\noindent{\emph{Inclined circular vortices}}
 \vskip 2mm
\noindent 
 \begin{figure}
\vskip 0mm
(a)\includegraphics[width=0.5\textwidth, trim=20mm 80mm 0mm 90mm]{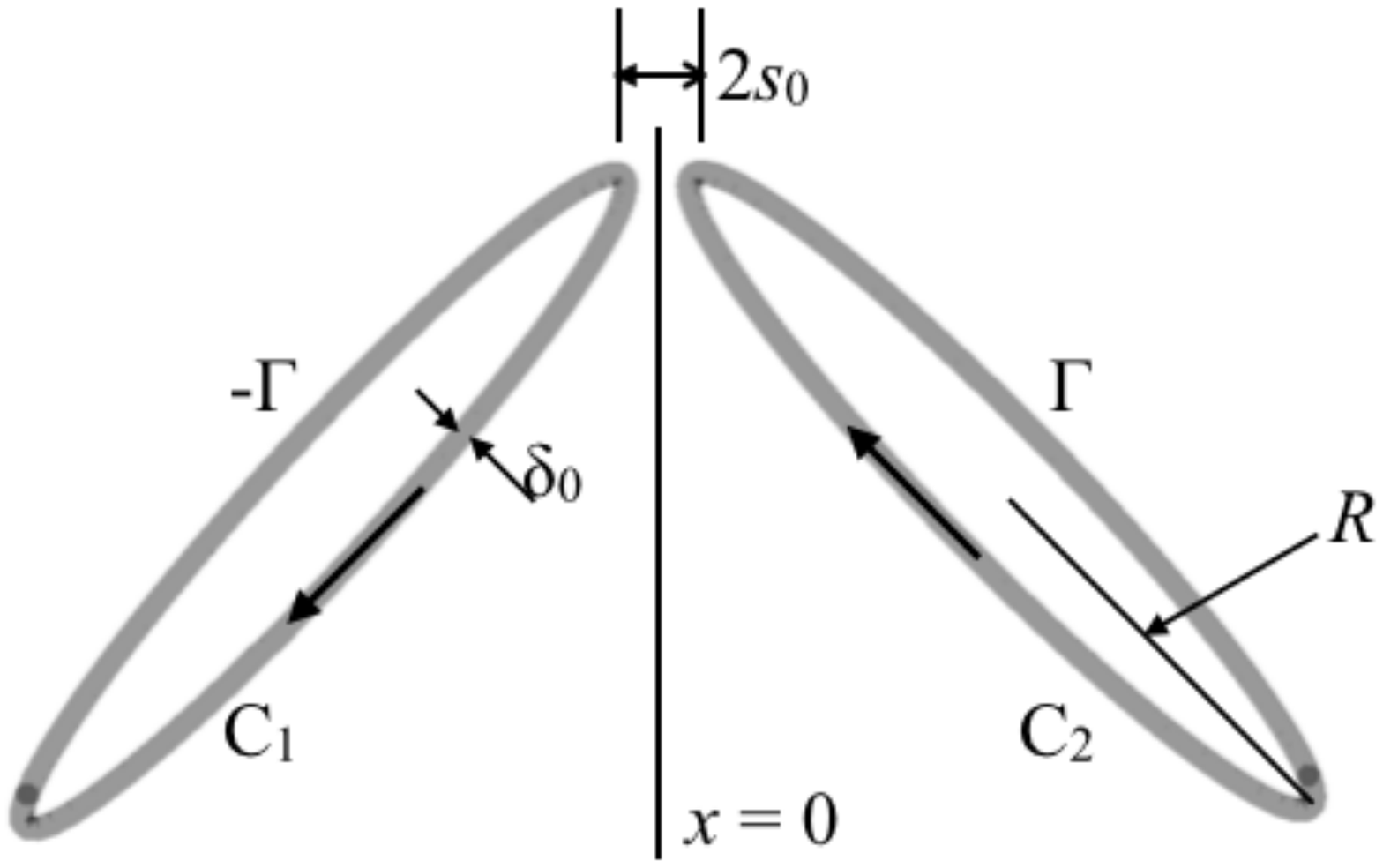}\\
\vskip -5mm
(b)\includegraphics[width=0.5\textwidth, trim=0mm 0mm 0mm 0mm]{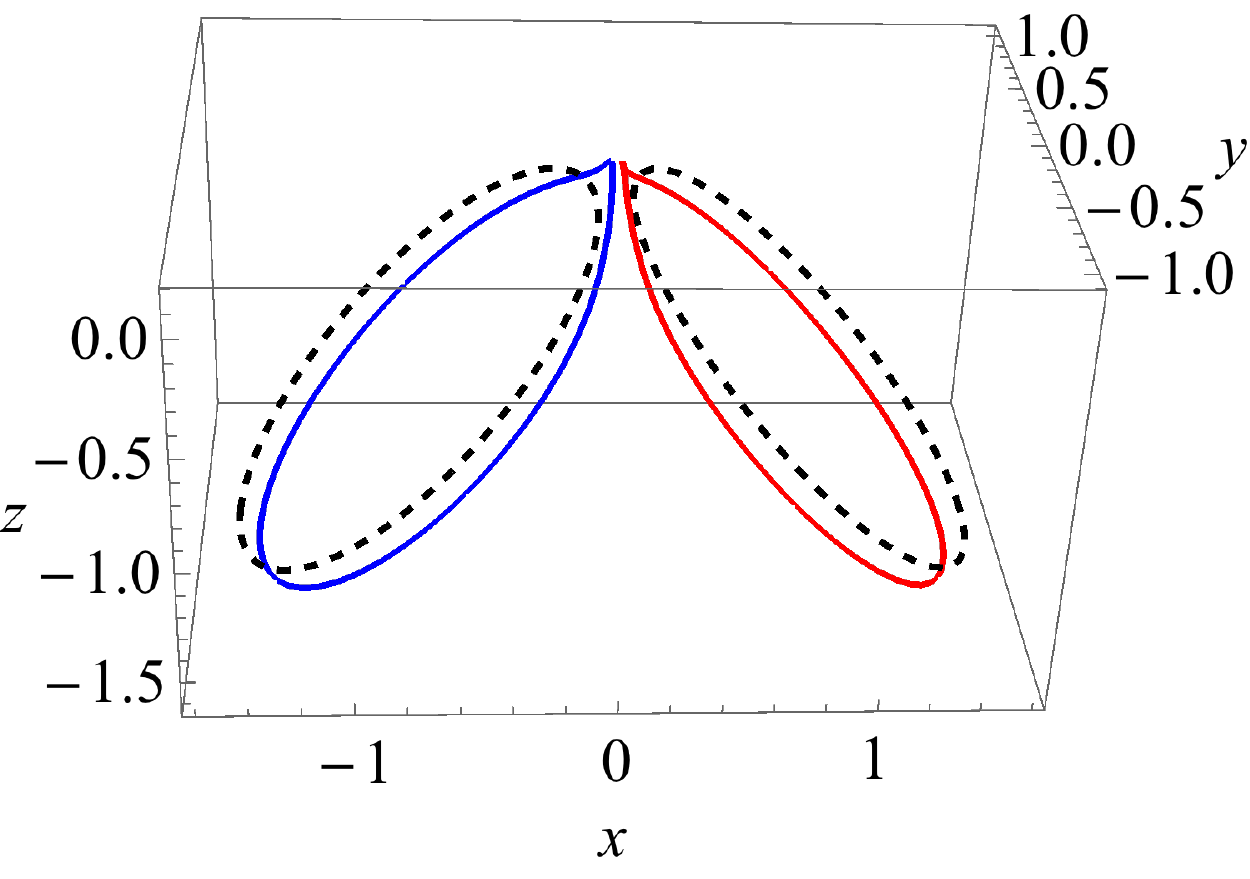}
\caption{(a) Two circular vortices of radius $R=\kappa_{0}^{-1}$ propagate towards each other at angle $2\alpha$; the arrows indicate the direction of vorticity $\bom$; the planes $x=0$ and $y=0$ are planes of symmetry; the vortices are assumed to have Gaussian core cross-sections of scale $\delta_{0}$, the minimum separation is $2s_{0}$, and it is assumed that  $\delta_{0}\ll s_{0}\ll R$; (b) early deformation of the  vortices, the dashed curves indicating the positions at $t=0$; although the vortices on the whole propagate downwards, the tipping points (i.e.~the points of nearest approach) move upwards and towards the plane $x=0$. [After Moffatt \& Kimura 2019a.]}
        \label{Fig_sketch}
 \end{figure}
Our starting point was to consider as an initial condition two  vortices  of circulations $\mp \Gamma$ centred on circles C$_1$ and C$_2$ of equal radii $R=\kappa_{0}^{-1}$ placed symmetrically on planes inclined to the plane $x=0$ at angles $\pm\alpha$, and oriented so that they propagate towards each other, as sketched in Fig.~\ref {Fig_sketch}a.  The minimum separation of the vortices is $2s_{0}$ and they are assumed to have Gaussian cross-section of scale $\delta_{0}$ where, by assumption,
\be
\delta_{0}\ll s_{0}\ll \kappa_{0}^{-1}.
\ee
Moreover, it is assumed that the vortex Reynolds number is large:
\be
R_{\Gamma}\equiv \Gamma/\nu \gg 1.
\ee
Under these conditions, the initial kinetic energy of the flow is finite, and for $t>0$ this energy decreases as a result of viscous dissipation.  

Let $\tn{T}_1$ and T$_2$ (the `tipping points') be the points of nearest approach of the two vortices.  For $t > 0$, the vortices (solid, blue and red online) deform in the manner indicated in Fig. \ref{Fig_sketch}(b).
What is most evident here is that the separation of the tipping points $2s(t)$ decreases and the curvature at these points $\kappa(t)$ increases.   The two vortices obviously do not remain circular; however, in a local description, it is only the variables $s(t)$ and $\kappa(t)$ that matter, and the rate of change of these variables at any instant is presumably the same as it would be for vortices coinciding with the circles of curvature at T$_1$ and T$_2$  (and so with the current values of $s(t)$ and $\kappa(t)$).  On this basis, the dimensionless equations describing this behaviour, as derived by Moffatt \& Kimura (2019a), are
\be \label{system1}
\frac{\tn{d} s}{\tn{d}{ \tau}}=-\frac{\kappa\cos\alpha}{4\pi}\left[\log\left(\frac{s}{\delta}\right)+\beta_{1}\right] ,\quad \frac{\tn{d} \kappa}{\tn{d}{\tau}}=\frac{\kappa\cos\alpha\sin\alpha}{4\pi {s}^2}\,,
\ee
where $\tau=(\Gamma/R^{2})t$ is dimensionless time, and lengths are made dimensionless relative to $R$; $\delta(\tau)$ is the vortex-core radial scale at the tipping points, and $\beta_1= 0.4417$  for the assumed Gaussian core structure.   These equations, which are valid for so long as $\kappa s \ll 1$ and $\delta/s \ll 1$, are obtained by calculating the velocity field in the neighbourhood of T$_1$, say, taking account of the self-induced velocity of C$_1$ and the velocity induced by the vortex C$_2$. The details of this derivation are complicated, and need not be repeated here.

To close the system (\ref{system1}), an equation describing the variation of  $\delta(\tau)$ is also required.  This involves finding the rate of stretching, $\lambda(\tau)$, of either vortex at its tipping 
point. The result is $\lambda\sim \kappa \cos\alpha/4\pi s $, and the resulting asymptotic equation for $\delta$ may be written
\be \label{system1delta}
 \frac{\tn{d} \,\delta^2}{\tn{d}{\tau}}=\epsilon -\lambda(\tau) \,\delta^{2}=\epsilon -\frac{\kappa\cos\alpha}{4\pi {s}} \,\delta^{2}\,,
\ee
where $\epsilon=R_{\Gamma}^{-1} \ll 1$. [Note that this is just as for a Burgers stretched vortex, which would be steady if $\lambda$ were constant and $\delta=(\epsilon/\lambda)^{1/2}$.]  Here,  $\epsilon$ in (\ref{system1delta}) is just the expected term representing viscous diffusion of the vortex core, while the term $-(\kappa\cos\alpha/4\pi s) \delta^{2}(\tau)$ represents stretching of the vortex, which reduces its core radius; this latter term becomes increasingly important and ultimately dominant as $\kappa$ increases and $s$ decreases.

The rate-of-strain tensor $e_{ij}$ can also be derived analytically; its principal axes in the neighbourhood of T$_1$ in the plane $y=0$ are always in the directions parallel to $x=\pm z$, with $x=+z$ being the direction of positive stretching.  This means that there is always a tendency to rotate the plane of the vortex near the tip towards the direction $\alpha=\pi/4$ (an effect that is masked in 
Fig.~\ref{Fig_sketch}(b) by the upward movement of the tipping points).  For this reason we adopted the constant value $\alpha=\pi/4$ in our computations. 

This same rate of strain tends to deform the Gauss\-ian vortex core from circular to elliptic form, the major axis of the ellipse being rotated anti-clockwise through $\pi/4$ relative to the positive strain direction, and so settling in the $z$-direction. This deformation is expected to be small when $R_{\Gamma}\gg 1$ (Moffatt, Kida \& Ohkitani 1992), essentially because the vortex then rotates so rapidly that it `feels' only the average inward rate of strain, this being compensated by the positive vortex stretching in the $y$-direction.  The assumption that the vortex core remains compact throughout the interaction process is debatable (McKeown et al. 2018, Kerr 2018, Hussain \& Duraisami 2011), and invites DNS investigation for the particular configuration (vortices initially on inclined planes) considered here.
\vskip 2mm
\noindent \emph{Partial Leray scaling}
\vskip 2mm
\noindent If for the moment we ignore the variation of the logarithmic term $\Delta=\log{s/\delta}+\beta_1$ in (\ref{system1}), so that $\Delta\approx\Delta_0$, then this pair of equations admits self similar solutions of the form
\be\label{scalings_Leray}
s^{2}(\tau)=s_{0}^{2}(1-\tau/\tau_{c})\quad \tn{and}\quad\kappa^{-2}(\tau)=\kappa_{0}^{-2}(1-\tau/\tau_{c})\,,
\ee
where $\tau_{c}\approx 4\pi s_{0}^2/\sin 2\alpha$.
Thus, the `singularity time' $\tau_c$ is determined by this sol\-ution; moreover,
this `Leray scaling' requires that $\Delta_0=\sin\alpha/s_0$, i.e.
\be\label{in_cond_Leray2}
\delta_{0}/s_{0}=\exp\left[\beta_{1}-\sin\alpha/s_{0}\right]\,,
\ee
so that the initial core radius is also determined if this Leray scaling is to be realised. If, for example, $s_{0}=0.1$, then, with $\alpha=\pi/2$, the requirement is that $\delta_{0}/s_{0}\approx1.3 \times 10^{-3}$.

This Leray scaling is `partial' because $\delta(\tau)$ does not conform to it. Indeed substituting (\ref{scalings_Leray}) in (\ref{system1delta}) gives
\be \label{system1delta2}
 \frac{\tn{d} \,\delta^2}{\tn{d}{\tau}}=\epsilon -\frac{\mu}{\tau_{c}-\tau} \,\delta^{2}\,,
\ee
where $\mu= s_{0}/2\sin \alpha$. The solution of (\ref{system1delta2}) has the behaviour $\delta^2 \sim (1-\tau/\tau_{c})^{\mu}$ as $\tau\rightarrow\tau_c$, and, since $\mu \ll 1$, this `plunges' to zero, with infinite negative gradient at ${\tau=\tau_{c}}$.
 \begin{figure}
\vskip 2 mm
\includegraphics[width=0.6\textwidth, trim=30mm 110mm 0mm 110mm]{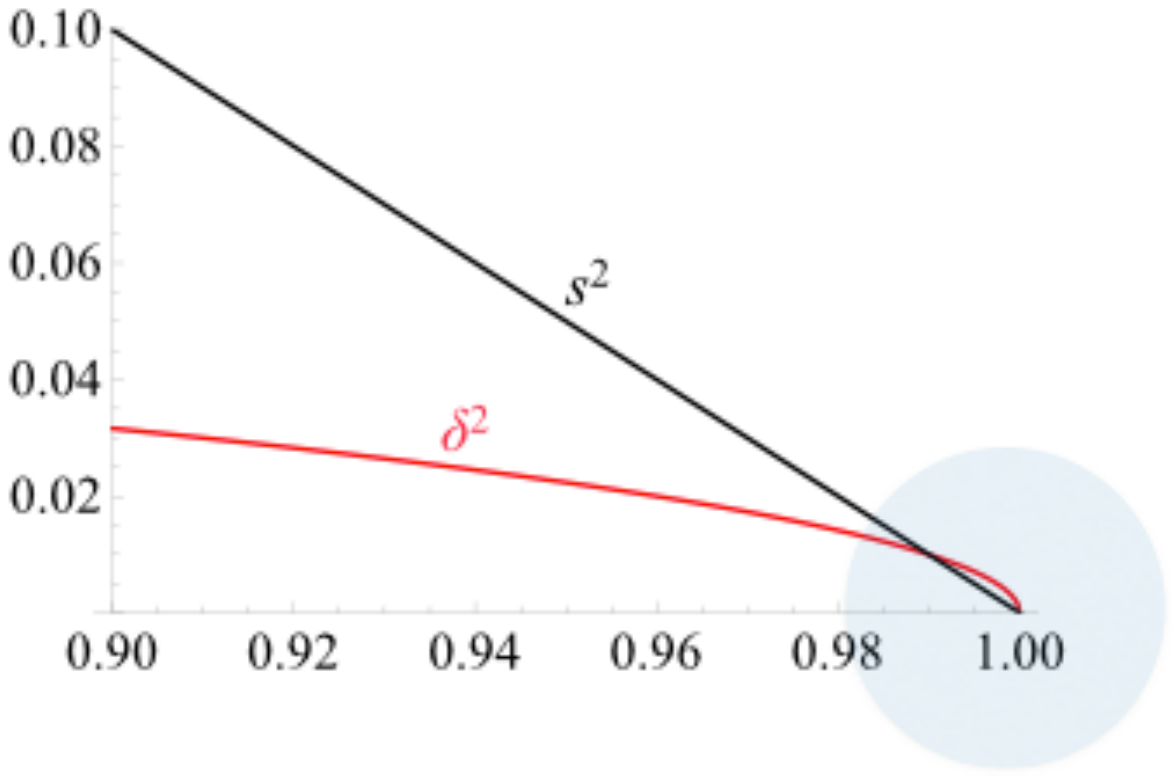}\\
\vskip 2mm
\caption{Sketch indicating that, no matter how small $\delta^{2}/s^{2}$ may be initially, there is always a neighbourhood of the singularity time $\tau_{c}$ where $\delta^{2}/s^{2} =\tn{O}(1)$; at this stage (indicated by the shaded circle) the vortices begin to overlap and viscous reconnection must be taken into account.}
        \label{Fig_05_near_singularity}
 \end{figure}

At first sight, this looks like a promising scenario for a  singularity; however, consideration of 
Fig.~\ref{Fig_05_near_singularity} reveals a problem:  because $\delta^2$ plunges to zero while $s^2$ decreases only linearly to zero as $\tau\rightarrow\tau_{c}$, there must inevitably be a small neighbourhood of 
$\tau=\tau_{c}$ (shaded in Fig.~\ref{Fig_05_near_singularity}) where $\delta^{2}/s^{2} =\tn{O}(1)$,  no matter how small the initial value  
$\delta_{0}^{2}/s_{0}^{2}$ may be.  The reason is essentially that the rate of decrease of $\delta^2$ over most of the time interval $0<\tau<\tau_c$ is so small that eventually $s^2$ decreases to O$(\delta^2)$; at this stage the two vortices begin to overlap in such a way that viscous reconnection becomes inevitable.  This requires the following  modification of the model to take account of this reconnection process.

\begin{figure}
\vskip 2 mm
\includegraphics[width=0.48\textwidth, trim=0mm 80mm 0mm 90mm]{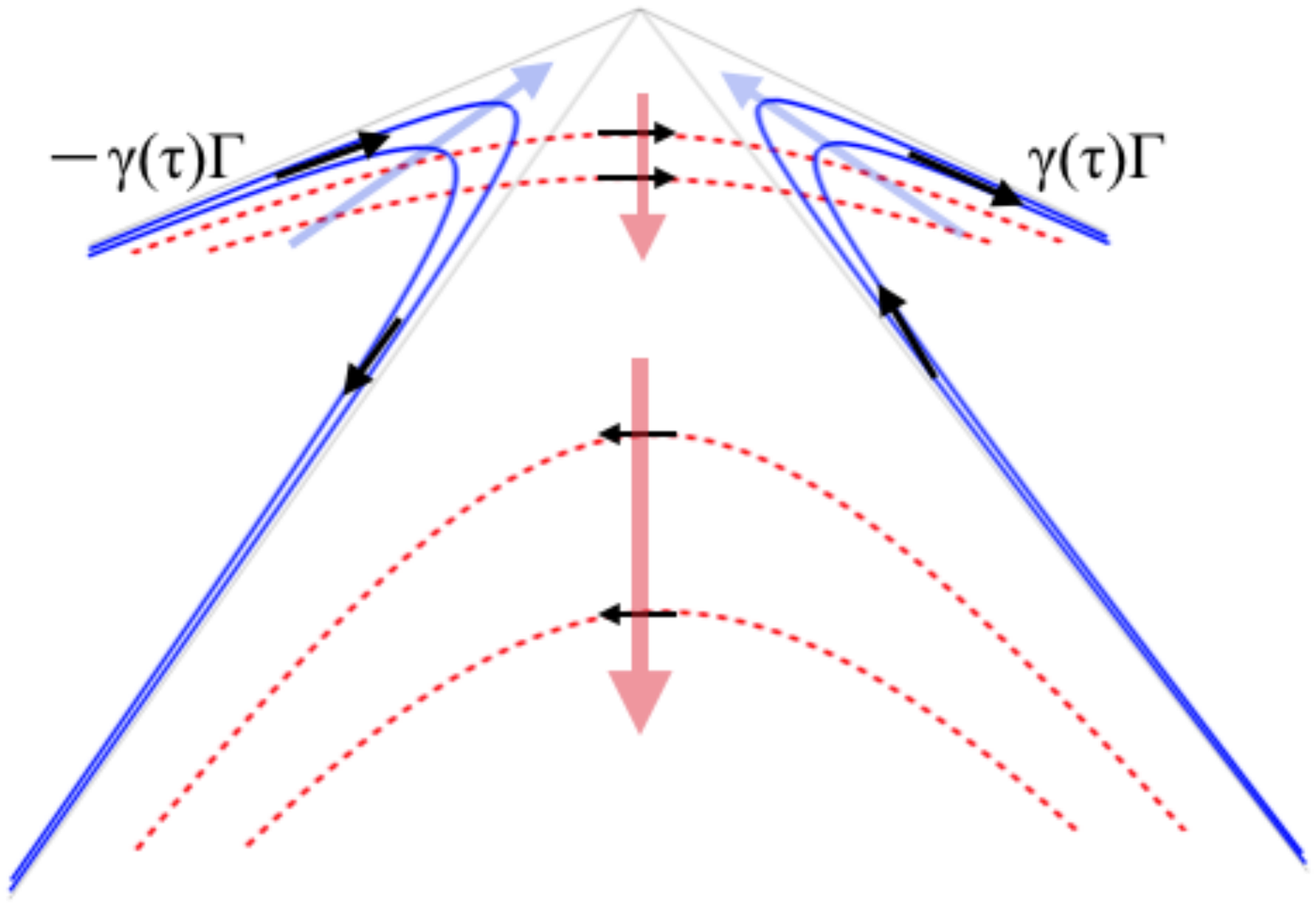}\\
\vskip 2mm
\caption{Sketch indicating the pyramid reconnection process very near to the apex of the pyramid; the initial circulations $\pm\Gamma$ split into reconnected circulations $\pm\Gamma_{r}$ (red) that propagate away from the apex and surviving circulations  $\pm\Gamma_{s}\equiv \pm\gamma(\tau)\Gamma$ (blue) that continue to propagate towards the apex; black arrows indicate the direction of vorticity, and coloured arrows indicate the direction of propagation. [After Moffatt \& Kimura 2019b.]}
        \label{Fig_06_pyramid_reconnection}
 \end{figure}
 \vskip 1mm
\noindent{\emph{The  reconnection phase}}
 \vskip 1mm
\noindent The situation is illustrated by the sketch of Fig.~\ref{Fig_06_pyramid_reconnection}.  The initial  circulations $\pm\Gamma$ split into reconnected circulations $\pm\Gamma_{r}$ (red) that propagate rapidly downwards away from the apex of the pyramid, and surviving circulations  $\pm\Gamma_{s}\equiv\pm\gamma(\tau)\Gamma$ (blue) that continue to propagate towards the apex. We expect $\gamma(\tau)$ to be monotonic decreasing from its initial value $\gamma(0)=1$ as reconnection proceeds. A similar splitting of circulation at the beginning of reconnection has been observed in DNS (Kerr 2018).

Reconnection occurs on the symmetry plane $x=0$, and the rate of reconnection can be calculated explicitly (Moffatt \& Kimura 2019b) on the assumption that the vortex cores remain compact during the reconnection process.   It is then found that $\gamma(\tau)$ evolves according to the equation
\be\label{reconnection}
\frac{\tn{d}\gamma}{\tn{d}\tau}=-\epsilon \frac{s\gamma}{2\sqrt{\pi}\delta^3}\exp{[-s^{2}/4\delta^2]}\,.
\ee
This equation makes good sense:  reconnection, represented by reduction of $\gamma(\tau)$, is evidently significant only when $s^2$ has decreased to O$(\delta^2)$.  

Eqn.~(\ref{reconnection}) must now be coupled with  (\ref{system1}) and (\ref{system1delta}) modified to take account of the reduced circulation, i.e.
\be \label{system1m}
\frac{\tn{d} s}{\tn{d}{ \tau}}=-\gamma\frac{\kappa\cos\alpha}{4\pi}\left[\log\left(\frac{s}{\delta}\right)+\beta_{1}\right] ,\quad \frac{\tn{d} \kappa}{\tn{d}{\tau}}=\gamma\frac{\kappa\cos\alpha\sin\alpha}{4\pi {s}^2}\,,
\ee
\vskip -2mm
\noindent and 
\vskip -6mm
\be \label{system1deltam}
 \frac{\tn{d} \,\delta^2}{\tn{d}{\tau}}=\epsilon -\gamma\frac{\kappa\cos\alpha}{4\pi {s}} \,\delta^{2}\,.
\ee
Eqns.~(\ref{reconnection}), (\ref{system1m}) and (\ref{system1deltam}) now constitute a fourth-order dynamical system which may be integrated numerically. Note that viscosity (via the parameter $\epsilon=\nu/\Gamma$) now appears in two places: in (\ref{reconnection}) representing the reconnection process, and in (\ref{system1deltam}) representing the natural viscous diffusion of either vortex on its own.

\begin{figure}
\vskip 2 mm
\includegraphics[width=0.45\textwidth, trim=0mm 30mm 0mm 30mm]{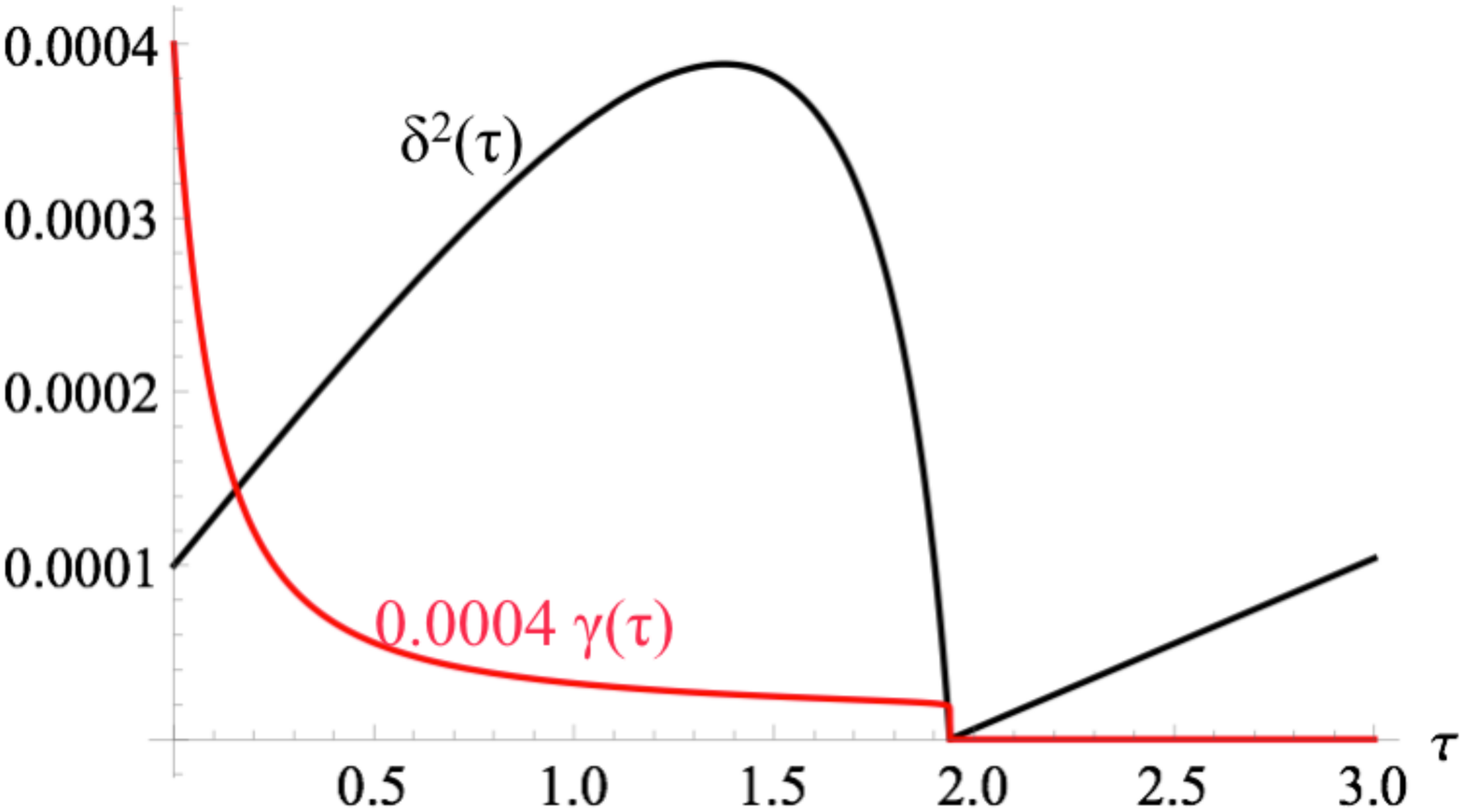}\\
\vskip 2mm
\caption{Evolution of $\delta^{2}(\tau)$ and $\gamma(\tau)$, for initial conditions $s(0)=0.1, \delta(0)=0.01$, $\kappa(0)=1,\gamma(0)=1$ and for $\epsilon^{-1}=3000$; the initial rise of $\delta^{2}(\tau)$ is due to viscous diffusion; the subsequent fall occurs when vortex stretching becomes dominant; $\delta^{2}(\tau)$ appears to fall to zero, but in fact attains a positive minimum $\delta_{\tn{min}}^2=2.727803\times10^{-24}$ as described in the text; the subsequent rise is again due to viscous diffusion acting on residual un-reconnected vorticity. [From Moffatt \& Kimura 2019b.]}
        \label{Fig_07_delta_to_zero}
 \end{figure}
  \begin{figure}
\vskip 2 mm
(a)\includegraphics[width=0.38\textwidth, trim=0mm 70mm 0mm 90mm]{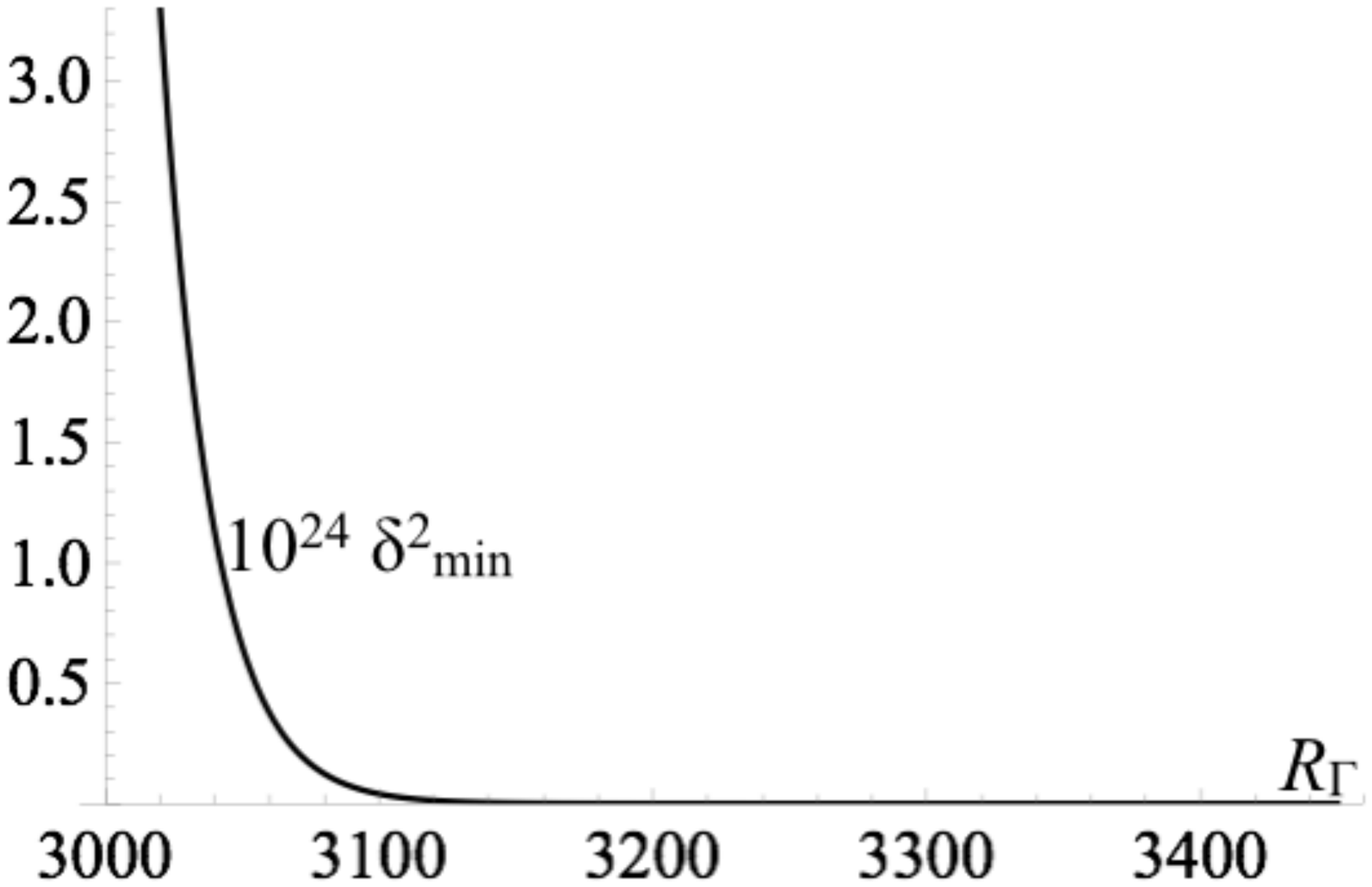}\\
\vskip 2mm
(b)\includegraphics[width=0.38\textwidth, trim=0mm 70mm 0mm 90mm]{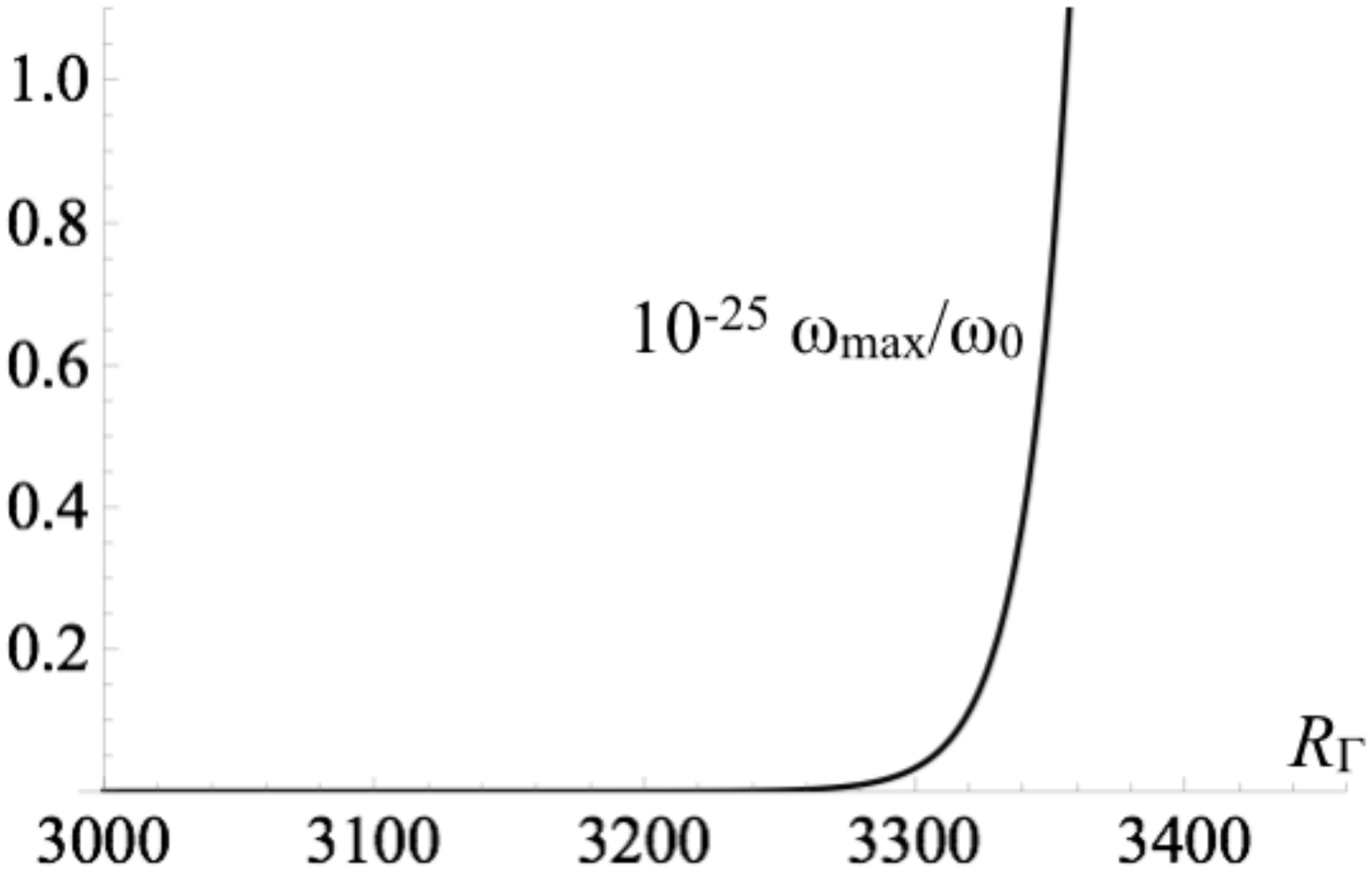}
        \caption{(a) The function (\ref{deltamin}) showing the extremely rapid decrease of $\delta_{\tn{min}}^{2}$ with increasing $R_\Gamma$;  (b) the function (\ref{omegamax}) showing the corresponding sharp increase of $\omega_{\tn{max}}$.[After Moffatt \& Kimura 2019b.]}
        \label{Fig_min_max}
 \end{figure}
The `surviving' vorticity (i.e.~the unreconnected vorticity, blue in Fig.~\ref{Fig_06_pyramid_reconnection}) is $\omega_{s}(\tau)=\gamma(\tau)/\delta^{2}(\tau)$ and although $\gamma(\tau)$ decreases, a singularity of $\omega_{s}$ may still conceivably occur;   a necessary condition for such a singularity is that there should exist a value of $\tau,\,\, \tau'_c$~say,  at which $\delta(\tau'_c)=0$  ($\tau'_c$ is distinct from $\tau_c$ because now the reconnection process is taken into account).  Fig.~\ref{Fig_07_delta_to_zero} shows the evolution of $\delta^{2}(\tau)$ for initial conditions $s(0)=0.1, \delta(0)=0.01$, $\kappa(0)=1,\gamma(0)=1$ and for $\epsilon=1/3000$.  The initial rise of $\delta^2$ is due to viscous diffusion; it  falls for $\tau \gtrsim 1.5$ when the stretching effect become dominant, and appears to fall to zero at about $\tau=1.94$.    However, close investigation shows that $\delta^2$ does not fall to zero, but attains a minimum value $\delta_{\tn{min}}^2=2.727803\times10^{-24}$ at 
$\tau=\tau'_c=1.93916197100167949177687$.  The minimum is extremely sharp, and $\tau'_c$ has to be determined to this degree of accuracy  to evaluate $\delta_{\tn{min}}^2$ correctly (we used Mathematica with 56-point precision). At this time $\gamma(\tau'_{c})=0.000451$, so that most of the vorticity has reconnected by this stage.  For $\tau>\tau'_c$, the residual un-reconnected vorticity is extremely weak, and expands due to viscous diffusion.

We carried out similar computations for values of $\epsilon^{-1}=R_\Gamma $ in the range $2000 < \epsilon^{-1}< 3400$, and found $\delta_{\tn{min}}^{2}(R_\Gamma)$ in each case. The results were consistent with a very steep decrease of  $\delta_{\tn{min}}^{2}(R_\Gamma)$ with increasing $R_\Gamma$ (Fig.~\ref{Fig_min_max}a), represented to good approximation by the formula
\be \label{deltamin}
 \delta^{2}_{\tn{min}}\sim \exp [-53(R_\Gamma /3000)^3]\,.
\ee
Although this is obviously extremely small for $R_\Gamma\gtrsim 3000$, it is not identically zero for any $R_\Gamma$, so the necessary condition for a strict mathematical finite-time singularity is not realised.

The corresponding maximum vorticity $\omega_{\tn{max}}$ occurs at a slightly earlier time $\tau_m (=1.9391619710016794917425$ when $\epsilon=1/3000$, differing from $\tau'_c$ only in the 20th decim\-al place); its dependence on  $R_\Gamma$ (Fig.~\ref{Fig_min_max}b) was found in the same range to be given by
\be\label{omegamax}
\omega_{\tn{max}}(R_\Gamma)/\omega_0 \sim  \exp [1 + 220(\log[R_\Gamma /2000])^2]\,,
\ee
where $\omega_0 = \gamma(0)/\delta(0)^2$ is its initial value, a behaviour more rapid than any power of $R_\Gamma$, and extremely large for $R_\Gamma\gtrsim 3000$. 
\section{Conclusions and Discussion}
The approach of Moffatt \& Kimura (2019a,b)  as summarised in \S V above,
provides  an analytical description of the evolution of two initially circular vortices symmetrically placed on inclined planes as in Fig.~\ref{Fig_sketch}a. The motion of each vortex due to the combination of its own self-induced velocity and the additional velocity induced by the other is calculated. The evolution of the radial scale $\delta$ of the core due to viscous diffusion and vortex stretching is incorporated, and a dynamical system involving $\delta(\tau)$,  (half) the minimal separation $s(\tau)$ and the curvature $\kappa(\tau)$ at the points of minimum separation is obtained, where $\tau$ is  dimensionless time.  This system exhibits Leray scaling for the variables $s$ and $\kappa$ implying singularity at a finite time $\tau_c$, but  $\delta$ does not share in this Leray scaling.  In fact the evolution is such that $\delta/s$, initially small, increases to O$(1)$ when $\tau$ approaches $\tau_c$, at which stage it is necessary to allow for vortex reconnection on the plane of symmetry. This is incorporated in terms of the variable $\gamma(\tau)=\Gamma_{s}/\Gamma$, where $\Gamma_{s}$ is the surviving (un-reconnected) vorticity.  The dynamical system is now fourth-order.  The variable $\delta(\tau)$ becomes extremely small when the vortex Reynolds number $R_\Gamma$ is large (eqn.~(\ref{deltamin}) and Fig.~\ref{Fig_min_max}(a)), but does not hit exactly zero for any value of $R_\Gamma$, so that a singularity is averted by the reconnection process.

In some respects, this phenomenon bears comparison with the free-surface cusp phenomenon described in \S\ref{free-surface_cusp}, in that, despite the smoothing effect of viscosity, the scale $\delta(\tau)$ becomes so small when $R_\Gamma\gtrsim 400$ that the continuum hypothesis on which the Navier-Stokes equation is based, is no longer tenable. In this sense, we are dealing here with a physical singularity, which is nevertheless mathematically non-singular.  In such circumstances, a  physical effect, so far neglected, must intervene to forestall the singular behaviour.  This physical effect depends on whether the fluid is liquid or gas: if  liquid, then cavitation is likely when the vorticity exceeds a critical value, and bubbles will form where these incipient singularities form; if gas, on the other hand, a pulse of sound may be emitted from the same small interaction region, much as described by Lighthill \&  Newman (1952).

\noindent\emph{Implications for turbulence}
\vskip 1mm
\noindent As already mentioned, turbulent flow at high Reynolds number is characterised by the fact that the mean rate of dissipation of energy $\Phi=\nu \langle {\bom}^2 \rangle$ is independent of $\nu$ in the limit $\nu\rightarrow 0$; somehow, the flow has to adapt so that energy supplied at a rate $\Phi$ on large scales is, in a statistically steady state, dissipated at this same rate, no matter how small $\nu$ may be. Over recent decades, experiments and DNS have increasingly revealed the presence of concentrated vortices within the flow; indeed, since Townsend's 1971 book \emph{The Structure of Turbulent Shear Flow}, turbulence has frequently and plausibly been regarded as a superposition of concentrated vortex structures. If vortex tubes  interact randomly in the manner described above, then energy dissipation will  be concentrated in the very localised regions of strong interaction.  This scenario is of course consistent with attempts to reconcile the phenomenon of intermittency with Kolmogorov's original 1941 theory (Frisch 1996).

In our representation of vortex interactions as described in \S V, the rate of dissipation of energy in one such reconnection event is $\Phi_{\tn{rec}}=\nu\int \bom_{s}^2 \tn{d}V\sim \nu \bom_{s\, \tn{max}}^2 \delta_{\tn{min}}^3 $, and since $\bom_{s \,\tn{max}}\sim \gamma /\delta_{\tn{min}}^2$, this leads to the dimensionless estimate
\be\label{reconn_diss}
\Phi_{\tn{rec}}\sim R_{\Gamma}^{-1}\gamma\,\delta_{\tn{min}}^{-1}\,.
\ee
Here $\gamma$ must be taken as the value of $\gamma(\tau)$ when $\omega_s$ is maximal, and is already small at this stage.  Even so, the large factor $\gamma \,\delta_{\tn{min}}^{-1}$ is amply sufficient to compensate the small factor $ R_{\Gamma}^{-1}$ in (\ref{reconn_diss}), so as to provide the required mechanism of energy dissipation. Each such reconnection event is extremely spiked in time as well as in space (rather like a spark), and therefore  difficult to detect in DNS;   we may nevertheless conjecture that, for any given turbulent Reynolds number, the fine-scale structure must evolve so that the number of such events that occur per unit volume and per unit time is just sufficient on average to dissipate the energy that  cascades down at the given rate $\Phi$ from the larger scales of the turbulence.

Here, we may discern an intriguing similarity with the soap-film collapse discussed in 
\S\ref{sec_twist_singularity}: in both cases, the near-singularity involves a topological transition (vortex reconnection in one case, a jump in surface topology in the other) that occurs at a finite time, dissipating a fin\-ite amount of energy by viscosity in a very small region of space and in a very small interval of time.  The analogy is no more than suggestive, but thought-provoking nonetheless.

\vskip 1mm
\begin{acknowledgments}
\noindent  It is a  pleasure to acknowledge collaborations with Ray Goldstein,  Adriana Pesci and Renzo Ricca on the work of \S III, with Jae-Tak Jeong on the work of \S IV, and  with Yoshi Kimura on the work of \S\S V and VI.
\end{acknowledgments}



\vskip 2mm
\begin{center}
\noindent {REFERENCES}
\end{center}
\vskip 2mm

\noindent Beale, J.~T., Kato,T. \& Majda, A.J 1984
Remarks on the breakdown of smooth solutions for the 3D Euler equations.
{\em Commun. Math. Phys.}, 94:61.
\vskip 2mm

\noindent Constantin, P,  Fefferman, C. \&  Majda A.J. 1996
 Geometric constraints on potentially singular solutions for the 3D Euler equations.
 {\em Comm. Partial Diff. Eqns.}, 21(3--4).
\vskip 2mm
\noindent  de~Waele, A.T.A.M. and  Aarts, R.G.K.M. 1994
Route to vortex reconnection.
{\em Phys. Rev. Lett.}, 72(4):482--485.
\vskip 2mm

\noindent Dean, W.~R.  \&  Montagnon, P.~E. 1949
On the steady motion of viscous liquid in a corner.
{\em Math. Proc. Cambridge Philos. Soc.}, 45(3):389--394.
\vskip 2mm

\noindent Eggers, J. 2001 Air entrainment through free-surface cusps. {\em Phys. Rev. Lett.} 86, 4290–4293.
\vskip 2mm

\noindent Frisch, U. 1996
 {\em Turbulence: The Legacy of A. N. Kolmogorov},
 Cambridge University Press.
\vskip 2mm

\noindent Goldstein, R.~E.,  McTavish, J., Moffatt, H.~K. \&  Pesci, A.~I. 2014
Boundary singularities produced by the motion of soap films.
{\em Proc. Natl. Acad. Sci.}, 111:8339--8344.
\vskip 2mm

\noindent Goldstein, R.~E.,  Moffatt, H.~K. \&  Pesci, A.~I. 2012
Topological constraints and their breakdown in dynamical evolution.
{\em Nonlinearity}, 25(10):R85--R98.
\vskip 2mm

\noindent Goldstein, R.~E.,  Moffatt, H.~K. \&  Pesci, A.~I. \&  Ricca, R.~L. 2010
Soap-film {M}\"{o}bius strip changes topology with a twist singularity.
{\em Proc. Natl. Acad. Sci.}, 107(51):21979--21984.
\vskip 2mm

\noindent Hussain, F. \& Duraisami, K. 2011
Mechanics of viscous vortex reconnection.
{\em Phys. Fluids}, 23:021701.
\vskip 2mm

\noindent  Jeong, J.~T. \&  Moffatt, H.~K. 1992
Free-surface cusps associated with flow at low {R}eynolds-number.
{\em J. Fluid Mech.}, 241:1--22.
\vskip 2mm

\noindent  Kerr, R.~M.
Enstrophy and circulation scaling for Navier-Stokes reconnection.
{\em J. Fluid Mech.}, 839:R2, 2018.
\vskip 2mm

\noindent Leray, J. 1934
Sur un liquide visqueux emplissant l'espace.
{\em Acta Mathematica}, 63:193--248.
\vskip 2mm

\noindent  Lighthill, M.~J. \&  Newman, M.~H.~A. 1952
On sound generated aerodynamically {I}. General theory.
{\em Proc. Roy. Soc. A}, 211:564--587.
\vskip 2mm

\noindent Van Dyke, M. 
{\em An Album of Fluid Motion}.
Parabolic Press, Stanford, 1982.
\vskip 2mm

\noindent McKeown, R.~, Ostilla-Monico, R.~,  Pumir, A.,  Brenner, M.P. \&  Rubinstein S.M. 2018
A cascade leading to the emergence of small structures in vortex ring collisions.
{\em Phys. Rev. Fluids}, 3:124702.
\vskip 2mm

\noindent Moffatt, H.~K. 1964
Viscous and resistive eddies near a sharp corner.
{\em J. Fluid Mech.}, 18(1):1--18.
\vskip 2mm

\noindent Moffatt, H.~K. , Goldstein, R.~E.  \&  Pesci, A.~I. 2016
 Soap-film dynamics and topological transitions under continuous deformation.
{\em Phys.Rev. Fluids}, 1(6):060503.
\vskip 2mm

\noindent  Moffatt, H.~K., Kida, S.~, \& Ohkitani, K. 1994
Stretched vortices -- the sinews of turbulence; large-Reynolds-number asymptotics.
{\em J. Fluid Mech.}, 259:241--264.
\vskip 2mm

\noindent Moffatt, H.~K.  \& Kimura, Y.  2019a
 Towards a finite-time singularity of the {N}avier-{S}tokes equations.
  {P}art 1. {D}erivation and analysis of dynamical system.
{\em J. Fluid Mech.}, 861:930--967.
\vskip 2mm

\noindent Moffatt, H.~K.  \& Kimura, Y.  2019b
 Towards a finite-time singularity of the {N}avier-{S}tokes equations.
 {P}art 2. {V}ortex reconnection and singularity evasion.
{\em J. Fluid Mech.}, 870:R1.
\vskip 2mm

\noindent Lord Rayleigh, 1920
\newblock {\em Scientific Papers}, 6:18.
\vskip 2mm

\noindent Shankar, P.~N. 2007
{\em Slow Viscous Flow}.
Imperial College Press.
\vskip 2mm

\noindent Siggia, E.~D. 1985
Collapse and amplification of a vortex filament.
{\em Phys. Fluids}, 28:794.
\vskip 2mm

\noindent Siggia, E.~D. \& Pumir,  A. 1985
Incipient singularities in the {N}avier-{S}tokes equations.
{\em Phys. Rev. Lett.}, 55:1749--1752.
\vskip 2mm

\noindent Taneda, S. 1979
Visualization of separating {S}tokes flows.
{\em J. Phys. Soc. Jpn.}, 46:1935--1942.
\vskip 2mm

\noindent Townsend, A.~A. 1971
{\em {The Structure of Turbulent Shear Flow (2nd edn.)}}.
Cambridge University Press.

\end{document}